\documentclass[12pt]{article}

\input amssym.def
\input amssym.tex

\begin{document}

\begin{titlepage}

\begin{flushright}
arXiv:1302.2598
\end{flushright}
\vskip 2.5cm

\begin{center}
{\Large \bf Neutrino Beam Constraints on\\
Flavor-Diagonal Lorentz Violation}
\end{center}

\vspace{1ex}

\begin{center}
{\large Brett Altschul\footnote{{\tt baltschu@physics.sc.edu}}}

\vspace{5mm}
{\sl Department of Physics and Astronomy} \\
{\sl University of South Carolina} \\
{\sl Columbia, SC 29208 USA} \\

\end{center}

\vspace{2.5ex}

\medskip

\centerline {\bf Abstract}

\bigskip

Breaking of isotropy and Lorentz boost invariance in the dynamics of 
sec\-ond-gen\-er\-a\-tion
leptons would lead to direction-dependent changes in the lifetimes of charged pions.
This would make the intensity of a neutrino beam produced via pion decay a function
of the beam orientation. The experimental signature of this phenomenon---sidereal
variations in the event rate at a downstream neutrino detector---has already been
studied, in searches for Lorentz-violating neutrino oscillations. Existing analyses
of MINOS near detector data can be used to constrain the flavor-diagonal
Lorentz violation coefficients affecting muon neutrino speeds at roughly the
$10^{-5}$ level.

\bigskip

\end{titlepage}

\newpage

\section{Introduction}

One exotic topic in elementary particle physics that has received a significant
amount of attention of late is the possibility that the fundamental laws of
nature may not be perfectly invariant under Lorentz and CPT
symmetries~\cite{ref-reviews}. Thus far,
there is no compelling experimental reason to doubt that these symmetries are exact.
However, there are important theoretical reasons to be interested in the topic.
Many prospective theories of quantum gravity may involve the breaking of these
symmetries, at least in certain regimes. The mechanisms for the symmetry breaking
include spontaneous breaking in
string theory~\cite{ref-kost18,ref-kost19} and theories of vector
fields~\cite{ref-altschul5},
effects in loop quantum gravity~\cite{ref-gambini,ref-alfaro} and
non-commutative
geometry~\cite{ref-mocioiu,ref-carroll3}, Lorentz violation through
spacetime-varying couplings~\cite{ref-kost20,ref-ferrero1}, and anomalous breaking of
Lorentz and CPT symmetries~\cite{ref-klinkhamer}.

The discovery of any kind of Lorentz or CPT violation would be extraordinarily
important---a sure sign of new physics, with a
fundamentally different structure from anything that has so far been
observed. Such a discovery would provide a remarkable window onto a fundamentally
new and different physical regime. This fact makes the subject area quite interesting,
for both theoretical and experimental research programs. Most recent theoretical work
is now performed within the context of effective quantum field theory.
There is an extensively-studied effective field theory
known as the standard model extension (SME) that contains all possible 
translation-invariant but Lorentz-violating operators that may be constructed out of
known standard model fields.

In the SME, the effects of Lorentz-violating operators are parameterized by small
tensor-valued coefficients~\cite{ref-kost1,ref-kost2}. In scenarios in which the
Lorentz violation arises spontaneously, these tensor coefficients represent
the nonzero vacuum expectation values of tensor-valued fields. Because CPT violation
in a quantum field theory that is both stable and unitary automatically entails the
existence of Lorentz violation as well~\cite{ref-greenberg}, the SME is automatically
the general effective field theory for describing CPT-violating physics.

As a systematic effective field theory, the SME contains, in principle, an infinite
number of possible Lorentz-violating operators. However, all but a small number of
these operators are expected to be suppressed at modest experimental energies. For
low-energy experiments, the most relevant subset of the general theory is the
minimal SME, which involves only those forms of Lorentz and CPT violation that are
gauge invariant, local, and superficially renormalizable. This minimal theory has become the standard framework used for parameterizing experimental Lorentz tests.
Recent searches for Lorentz violation have included studies of matter-antimatter
asymmetries for trapped charged
particles~\cite{ref-bluhm1,ref-gabirelse,ref-dehmelt1} and bound state
systems~\cite{ref-bluhm3,ref-phillips},
measurements of muon properties~\cite{ref-kost8,ref-hughes}, analyses of
the behavior of spin-polarized matter~\cite{ref-heckel3},
frequency standard comparisons~\cite{ref-berglund,ref-kost6,ref-bear,ref-wolf},
Michelson-Morley experiments with cryogenic
resonators~\cite{ref-muller3,ref-herrmann2,ref-herrmann3,ref-eisele,ref-muller2},
Doppler effect measurements~\cite{ref-saathoff,ref-lane1},
measurements of neutral meson
oscillations~\cite{ref-kost10,ref-kost7,ref-hsiung,ref-abe,
ref-link,ref-aubert}, polarization measurements on the light from cosmological
sources~\cite{ref-carroll2,ref-kost11,ref-kost21,ref-kost22},
high-energy astrophysical
tests~\cite{ref-stecker,ref-jacobson1,ref-altschul6,ref-altschul7,ref-klinkhamer2},
precision tests of gravity~\cite{ref-battat,ref-muller4}, and others.
The results of these experiments set constraints on the various SME coefficients, and
up-to-date information about
most of these constraints may be found in~\cite{ref-tables}.

The neutrino sector of the standard model is one of the most difficult to study, since
neutrinos only interact weakly. However, there have been a number of analyses of how
Lorentz violation could affect neutrino oscillation phenomena. The general
framework for studying SME-driven neutrino oscillations was given
in~\cite{ref-mewes1}.
Based on this work, there have been experimental analyses of the results from the
MINOS~\cite{ref-adamson1,ref-adamson2,ref-adamson3,ref-rebel},
LSND~\cite{ref-auerbach},
IceCube~\cite{ref-abbasi}, Double Chooz~\cite{ref-abe2}, and
MiniBooNE~\cite{ref-aguilar}
experiments. This paper will look at a different kind of
neutrino Lorentz violation, one not
involving oscillations; however, the ultimate physical observable is the same as
in some of the oscillation analyses. This means that elements drawn from previous
data analyses can immediately be carried over
into the present work.

Measuring flavor-diagonal Lorentz violation for neutrinos can be a challenge. The
MINOS analyses were intended as a search for off-diagonal SME
coefficients---for example, looking for oscillations $\nu_{\mu}\rightarrow\nu_{x}$
along a
baseline whose direction is changing slowly with the Earth's rotation. Particle
oscillation experiments can be extremely sensitive to small effects, because they
measure the quantum mechanically coherent build-up of tiny effects. In contrast,
the conventional way to identify neutrino Lorentz violation that does not involve
flavor-changing operators is to look for deviations from the expected neutrino
velocity $v_{\nu}=1$~\cite{ref-longo,ref-stodolsky,ref-adamson4,ref-opera}. Measuring
the velocities of neutral particles such as neutrinos is much more difficult than
equivalent measurements with charged species;
for charged leptons, it is possible to measure their velocities
indirectly, through their interactions with photons, because it is the four-velocity
that governs how a charge interacts with the electromagnetic field.

This paper will continue to focus on those forms of Lorentz violation that can modify
the neutrino velocities most pronouncedly. However, rather than direct measurements of
$v_{\nu}$, the focus will be on other effects generated by the same SME coefficients.
In particular, changes to the limiting neutrino velocity will also affect the rate at
which neutrinos are produced in weak decays. Most of the constraints described in this
paper concern coefficients affecting $\nu_{\mu}$ propagation that have never
previously been bounded. Despite a similarity in the notation (when flavor indices are
suppressed), these are not the same coefficients
studied in~\cite{ref-adamson1,ref-adamson3}; those results concerned coefficients
that were off-diagonal in flavor space and so could be constrained by looking at
flavor oscillation phenomena.

This paper is organized as follows. Section~\ref{sec-pion} will discuss how Lorentz
violation in the lepton sector affects the decay rates for charged pions.
Section~\ref{sec-minos} will discuss the influence of the specific experimental
geometry of the NuMI beam on pion decay observables.
New bounds based on previous analyses of the MINOS data are presented in
section~\ref{sec-bounds}, followed by a discussion of the outlook for future
experimental improvements.

\section{Lorentz Violation in Pion Decay}
\label{sec-pion}

Most of the SME coefficients for the first-generation fermions that make up
everyday matter have been measured quite precisely. However, forms of
Lorentz violation involving only second-generation particles are quite a bit more
difficult to constrain. Experiments that are sensitive to second-generation
Lorentz violation must involve either unstable charged particles or weakly coupled
neutrinos. Nonetheless, there are observables that depend rather sensitively on
the SME coefficients for the second-generation species. In particular, 
there are flavor-diagonal effects that could produce the same kind
of signature in the MINOS near detector as Lorentz-violating oscillations between
$\nu_{\mu}$ and another
species $\nu_{x}$. These phenomena cause the initial intensity of
the neutrino beam to depend on the direction in which the beam is pointing.
Periodic changes in the neutrino beam intensity would then change the
number of muon
charged current events recorded in the near detector, giving the same experimental
signature as Lorentz-violating neutrino oscillations.
However, comparison measurements using both the near and far detectors will
eliminate the dependence on the initial intensity of the neutrino beam; they can
therefore be used to distinguish changes to the neutrino speeds from Lorentz-violating
oscillation effects.

The most prominent reason that the NuMI beam intensity might vary with the sidereal
orientation is that Lorentz violation for the second-generation leptons
affects the lifetimes of the pions that decay to produce the beam. In any decay, the
dispersion relation for the daughter particles has a strong effect on the decay
rate, since it determines the amount of phase space available for the decay
products. Modifications of the muon and neutrino energy-momentum relations will
therefore change the rate at which pions decay to produce those particles. There
are also changes, of comparable size, to the weak interaction matrix element for the
process.

The details of the decay rate have been worked out~\cite{ref-altschul32}, to leading
order in the relevant Lorentz violation coefficients. The SME Lagrange density
relevant for the leptons in this scenario is
\begin{equation}
{\cal L}=i\bar{L}(\gamma^{\mu}+c_{L}^{\nu\mu}\gamma_{\nu})D_{\mu}L+
i\bar{R}(\gamma^{\mu}+c_{R}^{\nu\mu}\gamma_{\nu})D_{\mu}R\,+
{\rm mass\,terms},
\end{equation}
where $L$ and $R$ are the left- and right-chiral lepton multiplets,
\begin{equation}
L=\left[
\begin{array}{c}
\nu_{L} \\
\ell_{L}
\end{array}
\right],\,\,\,
R=\left[\ell_{R}\right].
\end{equation}
and $D_{\mu}$ is the covariant derivative, containing the electromagnetic and
weak gauge interactions. There are separate coefficients for the left-chiral and
right-chiral leptons; these can be alternately expressed in terms of Lorentz
violation coefficients $c^{\nu\mu}$ and $d^{\nu\mu}$, where
$c_{L}^{\nu\mu}=c^{\nu\mu}+d^{\nu\mu}$ and $c_{R}^{\nu\mu}=c^{\nu\mu}-d^{\nu\mu}$.
However, since only left-chiral particles undergo charged current weak interactions,
only $c_{L}$ affects the decay rate of charged pions.
%
%
The effects of flavor-diagonal minimal SME coefficients other than the $c$ and $d$ are
suppressed at relativistic energies.

Moreover, it is important to note that $SU(2)_{L}$ gauge invariance requires that the
Lorentz violation coefficients $(c_{L})^{\nu\mu}$ be the same for the left-chiral
charged
lepton and its corresponding neutrino. This means that it is not possible to have
neutrinos with Lorentz-violating dispersion relations without having the same kinds of
modifications also applying to the charged leptons---unless the chiral gauge
invariance (which is responsible for such features as lepton universality) is
abandoned.

The $c_{L}$ coefficients together form a traceless background tensor. The isotropic
element of this tensor, $(c_{L})_{00}$, endows spacetime with a preferred frame structure.
This leads to an isotropic violation of Lorentz boost symmetry. The other eight
independent coefficients also break spatial isotropy as well as boost invariance.
Although Lorentz violation is often accompanied by CPT violation, the effects of
the $c_{L}$ coefficients are all CPT-even.

The pion decay rate $\Gamma$ depends, at leading order, on the isotropic coefficient
$(c_{L})_{00}$, evaluated in the rest frame of the decaying pion. The dependence
is~\cite{ref-altschul32}
\begin{equation}
\label{eq-rate}
\Gamma=\Gamma_{0}\left[1+\frac{4}{1-m_{\mu}^{2}/m_{\pi}^{2}}(c_{L})_{00}\right],
\end{equation}
where $\Gamma_{0}$ is the rest frame decay rate in the absence of Lorentz violation.
The changes that lead to the modified $\Gamma_{0}$ are of two types. The first changes
are in the invariant matrix element for the process. A gauge field couples to the
average velocity of the incoming and outgoing particles at an interaction vertex; this
kind of velocity coupling is most familiar for electromagnetic interactions (in the
form of the classical Lorentz force law), but the weak coupling behaves similarly.
Having modified particle speeds therefore affects the pion decay matrix element.
Changes of the second type are kinematic in nature. Having unusual dispersion
relations for the outgoing particles changes the phase space available for the decay,
which in turn changes $\Gamma$. The phase space contributions are particularly
important in this case; they lead to the effects of $(c_{L})_{00}$ being
enhanced by the similarity in the pion and muon
masses. Using the physical values of these quantities, the decay rate is
$\Gamma=\Gamma_{0}\left[1+9.4(c_{L})_{00}\right]$.

In the pion rest frame, the total decay rate depends only on the $(c_{L})_{00}$ in
that frame. However, because decaying pions are typically moving relativistically
relative to the laboratory, $(c_{L})_{00}$ depends on the direction of that motion.
This gives the pion lifetime observed in the laboratory a sidereal dependence,
as the pion beam direction moves with the rotation of the Earth. Notably,
the sidereal dependence of the decay rate (and thus the neutrino beam strength) is
equally sensitive to violations of isotropy and violations of boost invariance. This
is a sharp contrast with the sensitivities seen in many low-energy laboratory
experiments, which are typically much more sensitive to anisotropy than boost
violation. The reason for the parity in sensitivity in this case is that the decaying
pions are moving relativistically in the laboratory, and this motion creates a strong
sensitivity to relativistic effects; as the Earthbound laboratory rotates, this
provides a way of testing whether the behavior of the pions is the same when they are
highly boosted in different directions.

Experimental
constraints on the SME coefficients are conventionally expressed using a set of
Sun-centered celestial equatorial coordinates~\cite{ref-bluhm4}.
The origin of these coordinates lies at the center of the Sun.
The $Z$-axis points parallel to the Earth's rotation axis; the $X$-axis
points toward the vernal equinox point on the celestial sphere; and the
$Y$-axis is  determined by the right hand rule. Time in these coordinates is $T$;
however, it
is generally advantageous to choose a local time coordinate $T_{\oplus}$ so that
at $T_{\oplus}=0$, the laboratory $y$-direction coincides with the $Y$-direction in
the Sun-centered frame.

The isotropic $(c_{L})_{00}$ in the pion rest frame is
\begin{equation}
\label{eq-c00}
(c_{L})_{00}=\gamma_{\pi}^{2}\left[(c_{L})_{TT}+(c_{L})_{(TJ)}(v_{\pi})_{J}+
(c_{L})_{JK}(v_{\pi})_{J}(v_{\pi})_{K}\right],
\end{equation}
in terms of the pion velocity $\vec{v}_{\pi}$, its Lorentz factor $\gamma_{\pi}$, and
the SME coefficients in the Sun-centered frame. The capital letter indices indicate
that the implied sums over the spatial coordinate indices $J$ and $K$ are being
performed in the Sun-centered frame.
The velocities associated with the
movement of the Earthbound laboratory are all nonrelativistic, so $\vec{v}_{\pi}$
is essentially the the pion's velocity relative to the lab.

When the particles involved in the decay are moving ultrarelativistically in the lab
frame, the relevant $(c_{L})_{00}$ effectively depends on the speed
$v_{\pi}\approx1$ only through the Lorentz factor $\gamma_{\pi}$.
Then the linear combination
$(c_{L})_{TT}+(c_{L})_{(TJ)}(\hat{v}_{\pi})_{J}+(c_{L})_{JK}(\hat{v}_{\pi})_{J}
(\hat{v}_{\pi})_{K}$, which depends only on the direction $\hat{v}_{\pi}$ of the
overall motion, becomes the key
quantity that determines the energy-momentum relation for ultrarelativistic leptons.
The dispersion relation for the highly
relativistic decay products (as observed in the lab frame) is
$E=[1-(c_{L})_{TT}-(c_{L})_{(TJ)}(\hat{v}_{\pi})_{J}-(c_{L})_{JK}(\hat{v}_{\pi})_{J}
(\hat{v}_{\pi})_{K}]|\vec{p}\,|$. This takes into account relativistic beaming, which
causes the decay products to be emitted along essentially the same direction as the
parent pion.

The dependence of $\Gamma$ on the pion beam direction is not the only manifestation
of anisotropy. In the pion rest frame, the decay is
actually anisotropic. Decay products are more likely to
emerge in some directions than others. In the pion frame, the rate for the decay channel with daughter particles emerging in directions $\hat{v}_{\mu}$ and
$\hat{v}_{\nu}=-\hat{v}_{\mu}$ may contain terms proportional to
$(c_{L})_{(TJ)}(\hat{v}_{\nu})_{J}$ or
$(c_{L})_{JK}(\hat{v}_{\nu})_{J}(\hat{v}_{\nu})_{K}$ at leading order. However,
in the total decay rate, which is found by integrating over all directions, all
the anisotropic terms cancel; the increased rates for decays into
certain directions compensate for decreased rates in other directions, leaving
the total decay rate only dependent on the lone isotropic term $(c_{L})_{00}$.
Moreover, the anisotropy of the decay products is difficult to
measure, because relativistic effects ensure that the decay products
are beamed into a
narrow pencil of angles around the direction $\hat{v}_{\pi}$ of the pion's motion.
It is worth noting that beaming is also responsible for the fact that it is
only the dispersion relation for particles moving essentially in
the $\hat{v}_{\pi}$-direction that determines the laboratory frame decay rate. That
this should be the case can be seen from the fact that if
the decay rate were (cumbersomely) evaluated in the lab frame, the phase space
available for the decay products would only depend on their energy-momentum relations
within their narrow range of allowed angles around $\hat{v}_{\pi}$.

While a full understanding of the decay anisotropy in the pion rest frame would be
fairly interesting, it is beyond the scope of the present analysis, since this
information is
not required in order to place order of magnitude bounds on the $c_{L}$ parameters
that are flavor diagonal (meaning those $c_{L}$ coefficients associated with
operators that do not mix multiple neutrino flavors).
In fact, calculating the angular dependence of the decay would
require a substantially more elaborate
generalization of the calculation in~\cite{ref-altschul32}. There are no
approximations apparent that could reliably be used to simplify such a calculation.
Phase space estimates are sometimes useful in evaluations of the rates for
Lorentz-violating processes. These estimates involve ignoring the effects of Lorentz
violation on the dynamical matrix elements involved and studying merely how
the Lorentz violation coefficients affect the phase space kinematics. This
strategy is typically useful when looking at processes that are normally forbidden
by Lorentz symmetry---that is, those for which the phase space available for the
process is conventionally zero~\cite{ref-stecker,ref-lehnert1,ref-altschul9}. It is
not surprising that in
such situations, the changes to the kinematics should predominate. However, the
calculations in~\cite{ref-altschul32} show that for pion decay, the effects of the
$c_{L}$ coefficients on the dynamics and kinematics are comparable in size; a phase
space estimate of the pion decay rate will not be quantitatively accurate.

Another simplification would apply if both the daughter particles were moving
ultrarelativistically
in the center-of-mass frame, as is the case for the $\pi^{+}\rightarrow e^{+}+
\nu_{e}$ decay mode. In such a scenario, there is no dynamical or kinematical
distinction between the two outgoing particles in the absence of Lorentz violation
[because of the Lorentz-invariant theory's $SU(2)_{L}$ and CP symmetries]. The equivalence of the daughter particles, which travel in antipodal directions in the
center-of-mass frame,
would preclude any anisotropy that had different rates of neutrino production in
the $\hat{p}$- and $-\hat{p}$-directions. This would rule out all odd multipole
patterns; in particular, there could be no dependence on the dipolar $(c_{L})_{(0j)}$
coefficients, and any anisotropy would have to come from the $(c_{L})_{jk}$.
Unfortunately, for the decay to a
muon-neutrino pair, the muon is not energetic enough for this simplification
to occur.

\section{Effects of Experimental Geometry}
\label{sec-minos}

In a neutrino oscillation experiment like MINOS, a detector intercepts the neutrino
beam near the beam's point of origin. If Lorentz violation in the neutrino sector
leads to sidereal oscillations in the strength of the neutrino beam,
the oscillations will be
detectable using the near detector. In this section, we will focus specifically
on understanding how to make such a measurement using
the NuMI neutrino beam and the MINOS detector located at Fermilab, because there
is more information available about possible sidereal oscillations for MINOS than
for other high-energy neutrino experiments.

The most elementary constraint based on the variability of the pion decay rate arises
from the fact that there can be a threshold boost beyond which the decay
$\pi^{+}\rightarrow\mu^{+}+\nu_{\mu}$ becomes energetically impossible. This effect
is nonperturbative in the $c_{L}$ coefficients, and so it was not included in the perturbative analysis that led to (\ref{eq-rate}). Nonetheless, if
$E_{\pi}>\sqrt{(m_{\pi}^{2}-m_{\mu}^{2})/2[-(c_{L})_{TT}-(c_{L})_{(TJ)}(v_{\pi})_{J}-
(c_{L})_{JK}(v_{\pi})_{J}(v_{\pi})_{K}]}$ (when the square root is real), the
greater-than-normal
growth of the muon and neutrino energies as functions of $|\vec{p}\,|$
means that there is not enough energy
to produce daughter particles with the required
momentum~\cite{ref-bi,ref-mestres}. The primary pion decay channel becomes disallowed.
(If the square root is imaginary, no such threshold exists; this occurs if the
neutrino velocity in the relevant direction is less than the speed of light, and so
the neutrino energy $E_{\nu}=v_{\nu}p_{\nu}$
grows more slowly than in the Lorentz-invariant theory.)

Naturally,
the existence of an upper threshold for the pion decay process could lead to
sidereal variations in the beam. The nonoccurrence of the decay above a
certain pion energy will affect the neutrino intensity. As the beam direction
rotates with the Earth, the threshold will vary, and a variable fraction of the
expected population of high energy neutrinos will fail to be produced. However, a
constraint derived from an analysis of this phenomenon appears to be
less sensitive than one that can be extracted from an analysis
of the decay rate based on the perturbative formula (\ref{eq-rate}).

The energy distribution of the neutrinos in the NuMI beam is somewhat complicated.
However, in order to get an order of magnitude constraint on the $c_{L}$ coefficients
for second-generation leptons, we shall treat the problem as the decay of a $\sim6$
GeV pion, without regard to the direction of the decay neutrino.
This energy is chosen to simulate a neutrino beam with a peak at $\sim 3$ GeV. 
A full analysis of the beam energy structure would introduce a number of additional
complications. The degree of forward beaming depends on the progenitor pion energy,
and this interacts nontrivially with the effects of Lorentz violation as well.
While the total decay rate $\Gamma$ depends only on the isotropic $(c_{L})_{00}$ in
the pion rest frame, the amplitudes for specific angular decay channels depend on the
anisotropic coefficients contracted with the momentum vectors of the daughter
particles. Certain neutrino production directions will be favored. Especially at
lower pion energies, where beaming is less pronounced, this can affect the
energy composition of the NuMI beam. In contrast, at high energies, the decay
neutrinos are strongly aligned into the observable beam, and the sensitivity to
Lorentz violation is simultaneously higher, because of a larger factor
$\gamma_{\pi}$ appearing in the boosted expression for $(c_{L})_{00}$.

Other experimental complications may be dealt with more explicitly.
The result (\ref{eq-rate}), in conjunction with the formula (\ref{eq-c00}) for
$(c_{L})_{00}$, gives the instantaneous decay rate for pions moving in a particular
direction. However, what is measured with an observation of the neutrino beam
strength is not the decay rate but the total number of pions that manage to
decay during the neutrino beam production process. This means that
the length of the pion decay pipe affects the magnitude of any variations in
the neutrino beam strength. If the pion decay rate, as measured in the laboratory
frame, is $\Gamma/\gamma_{\pi}$, then for pions moving at speed $v_{\pi}\approx1$
along a decay pipe of
length $D$, the fraction that decay over the length of the pipe is
$P(D)=1-e^{-\Gamma D/\gamma_{\pi}}$. The fractional difference in $P(D)$, if there is
a fractional change in the decay rate $\Delta \Gamma/\Gamma_{0}$, is
\begin{equation}
\label{eq-pipe}
\left(\left.\frac{1}{P}\frac{dP}{d\Gamma}\right|_{\Gamma=\Gamma_{0}}\right)\Delta
\Gamma=\left(\frac{\Delta \Gamma}{\Gamma_{0}}\right)\frac{\Gamma_{0}D/\gamma_{\pi}}
{e^{\Gamma_{0}D/\gamma_{\pi}}-1},
\end{equation}
with a factor of $\Gamma_{0}/\Gamma_{0}$ inserted to make the expression of product
of two dimensionless quantities.
Note that for $D\gg\gamma_{\pi}\Gamma_{0}^{-1}$, essentially all the pions have
time to decay, so
a change in the decay rate does not affect the total beam strength. In contrast, if
$D\ll\gamma_{\pi}\Gamma_{0}^{-1}$, the decaying fraction is
$P(D)\approx\Gamma D/\gamma_{\pi}$, and the
fractional change in the number of decays is simply $\Delta \Gamma/\Gamma_{0}$. As $D$
increases, the sensitivity of the beam intensity is suppressed; however, there is
also naturally an improvement in statistics at the detector, as the total number of
the neutrinos in the beam grows. The optimal sensitivity for a monoenergetic beam
will occur in the vicinity of $D\sim\gamma_{\pi}\Gamma_{0}^{-1}$.
For the NuMI beam, the decay pipe length is 677 m long, chosen to be comparable to
$\gamma_{\pi}\Gamma_{0}^{-1}=\gamma_{\pi}(2.6\times 10^{-8}\,{\rm s})$ for
pions with GeV energies. For a 6.0-GeV pion, the suppression factor
$\Gamma_{0}D/\gamma_{\pi}(e^{\Gamma_{0}D/\gamma_{\pi}}-1)$
is 0.31.

Finally, the most important influence that the specific geometry of the MINOS
experiment has on the possibility of sidereal oscillations in the detector event rate
comes from the details of how the pion decay direction changes with time. Let the
detector be located at colatitude $\chi$ ($\chi=42.18^{\circ}$ for the MINOS near
detector at Fermilab), and let $(\theta,\phi)$ be spherical coordinates describing
the angle between the beam direction and the local zenith direction
($\theta=93.27^{\circ}$) and the azimuthal angle in the plane of the Earth's surface,
measured starting eastward from south ($\phi=203.91^{\circ}$). Then the beam direction
at a local time $T_{\oplus}=0$ is~\cite{ref-mewes2}
\begin{eqnarray}
\hat{v}_{\pi} & = & N_{1}\hat{X}+N_{2}\hat{Y}+N_{3}\hat{Z} \\
& = & (\cos\chi\sin\theta\cos\phi+\sin\chi\cos\theta)\hat{X}
+(\sin\theta\sin\phi)\hat{Y} \nonumber\\
& & +(-\sin\chi\sin\theta\cos\phi+\cos\chi\cos\theta)
\hat{Z} \\
& = & -0.715\hat{X}-0.405\hat{Y}+0.571\hat{Z}.
\end{eqnarray}
As the Earth rotates with sidereal frequency $\omega_{\oplus}$, the key quantity
$(c_{L})_{00}$ varies according to
\begin{equation}
\label{eq-c00osc}
(c_{L})_{00}=\gamma_{\pi}^{2}\left[{\cal A}_{0}+{\cal A}_{\omega}\cos(\omega_{\oplus}
T_{\oplus})+{\cal B}_{\omega}\sin(\omega_{\oplus}T_{\oplus})+
{\cal A}_{2\omega}\cos(2\omega_{\oplus}T_{\oplus})+
{\cal B}_{2\omega}\sin(2\omega_{\oplus}T_{\oplus})\right],
\end{equation}
where the coefficients are
\begin{eqnarray}
{\cal A}_{0} & = & (c_{L})_{TT}+N_{3}(c_{L})_{(TZ)}
+N_{3}^{2}(c_{L})_{ZZ}+\frac{1}{2}(1-N_{3}^{2})[(c_{L})_{XX}+(c_{L})_{YY}] \\
{\cal A}_{\omega} & = & N_{1}(c_{L})_{(TX)}+N_{1}N_{3}(c_{L})_{(XZ)}+
N_{2}(c_{L})_{(TY)}+N_{2}N_{3}(c_{L})_{(YZ)} \\
{\cal B}_{\omega} & = & -N_{2}(c_{L})_{(TX)}-N_{2}N_{3}(c_{L})_{(XZ)}+
N_{1}(c_{L})_{(TY)}+N_{1}N_{3}(c_{L})_{(YZ)} \\
{\cal A}_{2\omega} & = & \frac{1}{2}(N_{1}^{2}-N_{2}^{2})(c_{L})_{-}
+N_{1}N_{2}(c_{L})_{(XY)} \\
{\cal B}_{2\omega} & = & -N_{1}N_{2}(c_{L})_{-}
+\frac{1}{2}(N_{1}^{2}-N_{2}^{2})(c_{L})_{(XY)},
\end{eqnarray}
with $(c_{L})_{-}=(c_{L})_{XX}-(c_{L})_{YY}$.

The ultimate experimental observable---the number of muon events recorded in the
detector---will therefore be modulated at the frequencies $\omega_{\oplus}$ and
$2\omega_{\oplus}$. Collecting all the factors to determine the sensitivity,
the fractional modulation in the normalized event rate will have amplitudes
equal to the ${\cal A}$ and ${\cal B}$ coefficients times a sensitivity factor
${\cal S}$. The sensitivity factor is assembled from several pieces. There is
the boost factor $\gamma_{\pi}^{2}$ [from (\ref{eq-c00}) and (\ref{eq-c00osc}], which
represents the enhancement of the boost invariance violation effects controlled by
$(c_{L})_{00}$ due to the relativistic motion of the center of mass. There is the
also $c_{L}$-dependent term from $\Gamma/\Gamma_{0}$, which tells how much
$(c_{L})_{00}$ affects the decay rate; and the result (\ref{eq-pipe}), which describes
how the modified decay rate $\Gamma$ in turn affects the beam intensity. These combine
to give
\begin{equation}
{\cal S}=\left(\gamma_{\pi}^{2}\right)
\left(\frac{4}{1-m_{\mu}^{2}/m_{\pi}^{2}}\right)
\left(\frac{\Gamma_{0}D/\gamma_{\pi}}{e^{\Gamma_{0}D/\gamma_{\pi}}-1}\right)
=5.4\times10^{3}.
\end{equation}
The largest part of the enhancement obviously derives from the highly relativistic
nature of the progenitor pions.

\section{Conclusions From MINOS Data}
\label{sec-bounds}

In order to constrain the $c_{L}$ coefficients for the second-generation leptons using
MINOS data, an analysis of the observed beam intensity versus sidereal time is
required. Fortunately, the core of the necessary analysis has already been performed
by the experimenters; this was the
basis of the papers~\cite{ref-adamson1,ref-adamson3}, the
conservative spirit of whose analyses
will be followed here. The observables for the scenario with
a direction-dependent rate of oscillations $\nu_{\mu}\rightarrow\nu_{x}$
and the scenario with a similarly direction-dependent pion decay rate are the same:
a change in the number of muon charged current events that indicate interactions of the beam neutrinos with the detector.
Since~\cite{ref-adamson1,ref-adamson3} found no significant evidence for sidereal
oscillations in the number of muon neutrinos in the beam, there can be no evidence of
a variation in the charged pion decay rate in these data sets either. Instead, there
will be constraints on the anisotropic $c_{L}$ parameters.

The MINOS analyses found no evidence of oscillations
that reached the $3\sigma$ level of significance. In fact,
the observed signal levels were well below such a level. The analysis
in~\cite{ref-adamson1} determined the Fast Fourier Transform power present in
each quadrature mode at frequencies $\omega_{\oplus}$ and $2\omega_{\oplus}$, as well
as additional higher harmonics. The level of statistical noise was characteristic of
a $1\sigma$ dispersion in these power values of $1.8\times10^{-2}$, when the data
were normalized to the total event rate (which was set by the rate of protons on
target in the collisions that originate the pions).
Thus a $3\sigma$ bound on the
normalized amplitudes ${\cal SA}$ and ${\cal SB}$ of the oscillations at the two
lowest frequencies is $5.4\times10^{-2}$. This corresponds to constraints on the
separate ${\cal A}$ and ${\cal B}$ coefficients of approximately $\times10^{-5}$.

The results of~\cite{ref-adamson3} are similar but used an antineutrino data set
with slightly less statistical power. Note that even in the presence of nonzero $d$
coefficients, the antineutrino data provide complementary constraints on the same
$c_{L}$ coefficients, which govern the dispersion relations for left-handed neutrinos
and their right-handed antiparticles.


Of the nine Lorentz-violating components of $(c_{L})^{\nu\mu}$, six can be constrained
by looking at sidereal variations in the event rates at multiple neutrino
experiments. The ones that cannot be constrained this way include $(c_{L})_{TT}$,
which is purely isotropic, as well as $(c_{L})_{(TZ)}$ and $(c_{L})_{ZZ}$, which
describe the effects of a preferred direction lying parallel to the Earth's axis.
In the spherical harmonic notation of~\cite{ref-mewes4}, the $c_{L}$ coefficients
affecting only a single generation of leptons are equivalent to the
$(c^{4}_{{\rm of}})_{jm}$, with highest weights up to $j=2$; the six
subject to constraint according to this method are those with $m\neq0$ in
celestial equatorial coordinates. For a single
particular experiment, four linear combinations of the remaining six parameters may
be constrained. These linear combinations are represented by the ${\cal A}$ and
${\cal B}$ coefficients.

\begin{table}
\begin{center}
\begin{tabular}{|c|c|}
\hline
Coefficient & Bound \\
\hline
$|(c_{L})_{(TX)}|$ & $2.3\times10^{-5}$ \\
$|(c_{L})_{(TY)}|$ & $2.3\times10^{-5}$ \\
$|(c_{L})_{-}|$ & $5.4\times10^{-5}$ \\
$|(c_{L})_{(XY)}|$ & $5.4\times10^{-5}$ \\
$|(c_{L})_{(XZ)}|$ & $4.0\times10^{-5}$ \\
$|(c_{L})_{(YZ)}|$ & $4.0\times10^{-5}$ \\
\hline
\end{tabular}
\caption{
\label{table-bounds}
Bounds on the magnitudes of the individual $c_{L}$ coefficients for second-generation
leptons, assuming only one coefficient at a time is nonvanishing.}
\end{center}
\end{table}

If the absence of sidereal oscillations in the signal is not due to a coincidental
cancellation among the coefficients, it is possible to set order of magnitude
constraints on the six separate $|(c_{L})_{(TX)}|$, $|(c_{L})_{(TY)}|$,
$|(c_{L})_{-}|$, $|(c_{L})_{(XY)}|$, $|(c_{L})_{(XZ)}|$ and $|(c_{L})_{(YZ)}|$
coefficients. This may be done by setting all but
one of the coefficients to zero, then finding the value of the remaining
coefficient that is required to raise one of the ${\cal A}$ or ${\cal B}$ to
an observable $3\sigma$ level. The results are given in
table~\ref{table-bounds}.

These order of magnitude constraints on the anisotropy coefficients,
at the $10^{-5}$ level, are
comparable to bounds on the time-averaged neutrino-light
velocity difference $v_{\nu}-1$ based on
terrestrial experiments with muon neutrinos. The sidereally averaged neutrino speed
measured in such experiments depends on the six constrained here and also on the
remaining coefficients $(c_{L})_{TT}$, $(c_{L})_{(TZ)}$, and $(c_{L})_{ZZ}$, as
outlined in~\cite{ref-altschul32}. Each such speed
measurement depends on a particular linear combination of the $c_{L}$
coefficients involved.

In future searches for anisotropic neutrino oscillations, based on muon neutrino
disappearance, it should be possible to add additional sensitivity to the
flavor-diagonal coefficients discussed here. The sensitivity to the $c_{L}$
coefficients
that affect solely second-generation leptons derives from the way the form of
Lorentz violation involved affect the decay of the pion progenitors of the beam
neutrinos.
The sensitivities of future measurements obviously depend most on the energies of the
pions involved; greater $\gamma_{\pi}$ values will generally improve constraints. The
position and orientation of the beam have lesser influences, as does the length of the
pion decay region. A very long decay pipe actually wipes out sensitivity in the
lower-energy component of the neutrino beam; the beam intensity is independent of
$c_{L}$ at a given energy if essentially all the pions in the relevant energy range
have sufficient time to decay. It is possible that this kind of shift toward higher
energies may actually be advantageous; although the signal is lost at lower energies,
the sensitivity to the $c_{L}$ coefficients increases with $\gamma_{\pi}^{2}$, and the
quality of the relativistic beaming approximation also increases with energy. A full
understanding of these effects will require more detailed understanding of the
parent pion and daughter neutrino beams' energy structures.

The current level of sensitivity is much lower than the sensitivity to
SME parameters that induce oscillations. It is also poorer than the $\sim10^{-11}$
sensitivity that has been obtained from the observed absence of the photon decay reaction $\gamma\rightarrow\mu^{-}+\mu^{+}$ for cosmic ray
photons~\cite{ref-altschul14}; however,
because the photon is a spin-1 state, the daughter particles in such
a reaction would necessarily have parallel spins, which makes such constraints
only sensitive to the spin-averaged $c$ coefficients, rather than the left-chiral
$c_{L}$. On the other hand, the sensitivity of the new method discussed in this paper
is comparable to that achieved for flavor-diagonal effects using direct time-of-flight
measurements, and this new strategy can represent an important tool for
constraining boost and isotropy violation in the difficult-to-access neutrino
sector.


\begin{thebibliography}{99}

\bibitem{ref-reviews}For overviews of recent work on Lorentz violation, see D.
Mattingly, Living Rev. Rel. {\bf 8}, 5 (2005) and the contents of {\em Proceedings of
the Fifth Meeting on {\rm CPT} and Lorentz Symmetry}, edited by V. A. Kosteleck\'{y}
(World Scientific, Singapore, 2011).
\bibitem{ref-kost18}V. A. Kosteleck\'{y}, S. Samuel, Phys. Rev. D {\bf 39}, 683
(1989).
\bibitem{ref-kost19}V. A. Kosteleck\'{y}, R. Potting, Nucl. Phys. B {\bf 359}, 545
(1991).
\bibitem{ref-altschul5}B. Altschul, V. A. Kosteleck\'{y}, Phys. Lett. B {\bf 628},
106 (2005).
\bibitem{ref-gambini}R. Gambini, J. Pullin, Phys. Rev. D {\bf 59}, 124021 (1999).
\bibitem{ref-alfaro}J. Alfaro, H. A. Morales-T\'{e}cotl, L. F. Urrutia, Phys. Rev.
D {\bf 65}, 103509 (2002).
\bibitem{ref-mocioiu}I. Mocioiu, M. Pospelov, R. Roiban, Phys. Lett. B {\bf 489},
390 (2000).
\bibitem{ref-carroll3}S. M. Carroll, J. A. Harvey, V. A. Kosteleck\'{y}, C. D.
Lane, T. Okamoto, Phys. Rev. Lett. {\bf 87}, 141601 (2001).
\bibitem{ref-kost20}V. A. Kosteleck\'{y}, R. Lehnert, M. J. Perry, Phys. Rev. D
{\bf 68}, 123511 (2003).
\bibitem{ref-ferrero1}A. Ferrero, B. Alt\-schul, Phys. Rev. D {\bf 80}, 125010 (2009).
\bibitem{ref-klinkhamer}F. R. Klinkhamer, C. Rupp, Phys. Rev. D {\bf 70}, 045020
(2004).
\bibitem{ref-kost1}D. Colladay, V. A. Kosteleck\'{y}, Phys. Rev. D {\bf 55},
6760 (1997).
\bibitem{ref-kost2}D. Colladay, V. A. Kosteleck\'{y}, Phys. Rev. D {\bf 58},
116002 (1998).
\bibitem{ref-greenberg}O. W. Greenberg, Phys. Rev. Lett. {\bf 89}, 231602 (2002).
\bibitem{ref-bluhm1}R. Bluhm, V. A. Kosteleck\'{y}, N. Russell, Phys. Rev.
Lett. {\bf 79}, 1432 (1997).
\bibitem{ref-gabirelse}G. Gabrielse, A. Khabbaz, D. S. Hall, C. Heimann, H.
Kalinowsky, W. Jhe, Phys. Rev. Lett. {\bf 82}, 3198 (1999).
\bibitem{ref-dehmelt1}H. Dehmelt, R. Mittleman, R. S. Van Dyck, Jr., P.
Schwinberg, Phys. Rev. Lett. {\bf 83}, 4694 (1999).
\bibitem{ref-bluhm3}R. Bluhm, V. A. Kosteleck\'{y}, N. Russell , Phys. Rev.
Lett. {\bf 82}, 2254 (1999).
\bibitem{ref-phillips}D. F. Phillips, M. A. Humphrey, E. M. Mattison, R. E.
Stoner, R. F. C. Vessot, R. L. Walsworth , Phys. Rev. D {\bf 63}, 111101(R)
(2001).
\bibitem{ref-kost8}R. Bluhm, V. A. Kosteleck\'{y}, C. D. Lane, Phys. Rev. Lett.
{\bf 84}, 1098 (2000).
\bibitem{ref-hughes}V. W. Hughes, {\em et al.}, Phys. Rev. Lett. {\bf 87},
111804 (2001).
\bibitem{ref-heckel3}B. R. Heckel, E. G. Adelberger, C. E. Cramer, T. S. Cook, S.
Schlamminger, U. Schmidt, Phys. Rev. D {\bf 78}, 092006 (2008).
\bibitem{ref-berglund}C. J. Berglund, L. R. Hunter, D. Krause, Jr., E. O.
Prigge, M. S. Ronfeldt, S. K. Lamoreaux, Phys. Rev. Lett. {\bf 75}, 1879 (1995).
\bibitem{ref-kost6}V. A. Kosteleck\'{y}, C. D. Lane, Phys. Rev. D {\bf 60},
116010 (1999).
\bibitem{ref-bear}D. Bear, R. E. Stoner, R. L. Walsworth, V. A. Kosteleck\'{y},
C. D. Lane, Phys. Rev. Lett. {\bf 85}, 5038 (2000).
\bibitem{ref-wolf}P. Wolf, F. Chapelet, S. Bize, A. Clairon,  Phys. Rev. Lett.
{\bf 96}, 060801 (2006).
\bibitem{ref-muller3}H. M\"{u}ller, {\em et al.}, Phys. Rev. Lett. {\bf 99}, 050401
(2007).
\bibitem{ref-herrmann2}S. Herrmann, A. Senger, K. M\"{o}hle, E. V. Kovalchuk, A.
Peters, in {\em CPT and Lorentz Symmetry IV}, edited by V. A. Kosteleck\'{y} (World
Scientific, Singapore, 2008), p. 9.
\bibitem{ref-herrmann3}S. Herrmann, {\em et al.}, Phys. Rev. D 80, 105011 (2009).
\bibitem{ref-eisele}Ch. Eisele, A. Yu. Nevsky, S. Schiller, Phys. Rev.
Lett. 103, 090401 (2009).
\bibitem{ref-muller2}H. M\"{u}ller, Phys. Rev. D {\bf 71}, 045004 (2005).
\bibitem{ref-saathoff}G. Saathoff, S. Karpuk, U. Eisenbarth, G. Huber, S. Krohn, R.
Mu\~{n}oz Horta, S. Reinhardt, D. Schwalm, A. Wolf, G. Gwinner, Phys. Rev. Lett.
{\bf 91}, 190403 (2003).
\bibitem{ref-lane1}C. D. Lane, Phys. Rev. D {\bf 72}, 016005 (2005).
\bibitem{ref-kost10}V. A. Kosteleck\'{y}, Phys. Rev. Lett. {\bf 80}, 1818
(1998).
\bibitem{ref-kost7}V. A. Kosteleck\'{y}, Phys. Rev. D {\bf 61}, 016002 (1999).
\bibitem{ref-hsiung}Y. B. Hsiung, Nucl. Phys. Proc. Suppl. {\bf 86}, 312
(2000).
\bibitem{ref-abe}K. Abe {\em et al.} (Belle Collaboration), Phys. Rev. Lett.
{\bf 86}, 3228 (2001).
\bibitem{ref-link}J. M. Link {\em et al.}, Phys. Lett. B {\bf 556}, 7 (2003). 
\bibitem{ref-aubert}B. Aubert {\em et al.} (BABAR Collaboration), Phys. Rev. Lett.
{\bf 96}, 251802 (2006).
\bibitem{ref-carroll2}S. M. Carroll, G. B. Field, Phys. Rev. Lett. {\bf 79},
2394 (1997).
\bibitem{ref-kost11}V. A. Kosteleck\'{y}, M. Mewes, Phys. Rev. Lett. {\bf 87},
251304 (2001).
\bibitem{ref-kost21}V. A. Kosteleck\'{y}, M. Mewes, Phys. Rev. Lett. {\bf 97},
140401 (2006).
\bibitem{ref-kost22}V. A. Kosteleck\'{y}, M. Mewes, Phys. Rev. Lett. {\bf 99}, 011601
(2007).
\bibitem{ref-stecker}F. W.  Stecker, S. L.  Glashow, Astropart. Phys. {\bf 16},  97
(2001).
\bibitem{ref-jacobson1}T. Jacobson, S. Liberati, D. Mattingly, Nature {\bf 424},
1019 (2003).
\bibitem{ref-altschul6}B. Altschul, Phys. Rev. Lett. {\bf 96}, 201101 (2006).
\bibitem{ref-altschul7}B. Altschul, Phys. Rev. D {\bf 74}, 083003 (2006).
\bibitem{ref-klinkhamer2}F. R. Klinkhamer, M. Risse, Phys. Rev. D {\bf 77}, 016002
(2008); addendum Phys. Rev. D {\bf 77}, 117901 (2008).
\bibitem{ref-battat}J. B. R. Battat, J. F. Chandler, C. W. Stubbs, Phys. Rev. Lett.
{\bf 99}, 241103 (2007).
\bibitem{ref-muller4}H. M\"{u}ller, S. W. Chiow, S. Herrmann, S. Chu, K.-Y. Chung,
Phys. Rev. Lett. {\bf 100}, 031101 (2008).
\bibitem{ref-tables}V. A. Kosteleck\'{y}, N. Russell, arXiv:0801.0287v6.
\bibitem{ref-mewes1}V. A. Kosteleck\'{y}, M. Mewes, Phys. Rev. D {\bf 69}, 016005
(2004).
\bibitem{ref-adamson1}P. Adamson, {\em et al.} (MINOS Collaboration), Phys. Rev.
Lett. {\bf 101}, 151601 (2008).
\bibitem{ref-adamson2}P. Adamson, {\em et al.} (MINOS Collaboration), Phys. Rev.
Lett. {\bf 105}, 151601 (2010).
\bibitem{ref-adamson3}P. Adamson, {\em et al.} (MINOS Collaboration), Phys. Rev. D
{\bf 85}, 031101(R) (2012).
\bibitem{ref-rebel}B. Rebel, S. Mufson, arXiv:1301.4684.
\bibitem{ref-auerbach}L. B. Auerbach, {\em et al.} (LSND Collaboration), Phys. Rev. D
{\bf 72}, 076004 (2005).
\bibitem{ref-abbasi}R. Abbasi, {\em et al.} (IceCube Collaboration), Phys. Rev. D
{\bf 82}, 112003 (2010).
\bibitem{ref-abe2}Y. Abe, {\em et al.} (Double Chooz Collaboration), Phys. Rev. D
{\bf 86}, 112009 (2012).
\bibitem{ref-aguilar}A. A. Aguilar-Arevalo, {\em et al.} (MiniBooNE Collaboration),
Phys. Lett. B {\bf 718} (2013).
\bibitem{ref-longo}M. J. Longo, Phys. Rev. D {\bf 36}, 3276 (1987). 
\bibitem{ref-stodolsky}L. Stodolsky, Phys. Lett. B {\bf 201}, 353 (1988).
\bibitem{ref-adamson4}P. Adamson, {\em et al.} (MINOS Collaboration), Phys. Rev. D
{\bf 76}, 072005 (2007).
\bibitem{ref-opera}T. Adam, {\em et al.}, JHEP {\bf 1210}, 093 (2012).
\bibitem{ref-altschul32}B. Altschul, Phys. Rev. D {\bf 84}, 091902(R) (2011).
\bibitem{ref-bluhm4}R. Bluhm, V. A. Kosteleck\'{y}, C. D. Lane, N. Russell, Phys.
Rev. D {\bf 68}, 125008 (2003).
\bibitem{ref-lehnert1}R. Lehnert, R. Potting, Phys. Rev. Lett. {\bf 93}, 110402
(2004).
\bibitem{ref-altschul9}B. Altschul, Phys. Rev. Lett. {\bf 98}, 041603 (2007).
\bibitem{ref-bi}X.-J. Bi, P.-F. Yin, Z.-H. Yu, Q. Yuan, Phys. Rev. Lett. {\bf 107},
241802, (2011).
\bibitem{ref-mestres}L. Gonzalez-Mestres, arXiv:1109.6630.
\bibitem{ref-mewes2}V. A. Kosteleck\'{y}, M. Mewes, Phys. Rev. D {\bf 70}, 076002
(2004).
\bibitem{ref-mewes4}V. A. Kosteleck\'{y}, M. Mewes, Phys. Rev. D {\bf 85}, 096005 (2012).
\bibitem{ref-altschul14}B. Altschul, Astropart. Phys. {\bf 28}, 380 (2007).

\end{thebibliography}
\end{document}